\documentclass[useAMS,usenatbib,usegraphicx]{mn2e}
\usepackage{txfonts}
\title[SXP6.85 undergoes type II outburst in the SMC]{Be/X-ray binary SXP6.85 undergoes large Type II outburst in the Small Magellanic Cloud}
\author[L. J. Townsend et al.]{L. J. Townsend$^{1}$\thanks{E-mail: ljt203@soton.ac.uk (LJT)}, M. J. Coe$^{1}$, V. A. McBride$^{1}$, A. J. Bird$^{1}$, M. P. E. Schurch$^{2}$, \newauthor R. H. D. Corbet$^{3}$, F. Haberl$^{4}$, J. L. Galache$^{5}$, A. Udalski$^{6}$\\
$^{1}$School of Physics and Astronomy, University of Southampton, Highfield, Southampton, SO17 1BJ, United Kingdom\\
$^{2}$Department of Astronomy, University of Cape Town, Private Bag X3, Rondebosch 7701, South Africa\\
$^{3}$University of Maryland Baltimore County, X-ray Astrophysics Laboratory, Mail Code 662, NASA Goddard Space Flight Center, Greenbelt, MD 20771, USA\\
$^{4}$Max-Planck-Institut f\"ur extraterrestrische Physik, Giessenbachstra\ss{}e, 85748, Germany\\
$^{5}$Harvard-Smithsonian Center for Astrophysics, 60 Garden St., Cambridge, MA 02138, USA\\
$^{6}$Warsaw University Observatory, Aleje Ujazdowskie 4, 00-478 Warsaw, Poland}

\begin{document}

\date{Accepted 2009 December 14.  Received 2009 December 14; in original form 2009 August 6}

\pagerange{\pageref{firstpage}--\pageref{lastpage}} \pubyear{2009}

\maketitle

\label{firstpage}

\begin{abstract}
The Small Magellanic Cloud (SMC) Be/X-ray binary pulsar SXP6.85 = XTE J0103-728 underwent a large Type II outburst beginning on 2008 August 10. The source was consistently seen for the following 20 weeks (MJD = 54688 - 54830). We present X-ray timing and spectroscopic analysis of the source as part of our ongoing \textit{Rossi X-ray Timing Explorer (RXTE)} monitoring campaign and \textit{INTEGRAL} key programme monitoring the SMC and 47 Tuc. A comparison with the Optical Gravitational Lensing Experiment (OGLE) III light curve of the Be counterpart shows the X-ray outbursts from this source coincide with times of optical maximum. We attribute this to the circumstellar disk increasing in size, causing mass accretion onto the neutron star. Ground based IR photometry and H$\alpha$ spectroscopy obtained during the outburst are used as a measure of the size of the circumstellar disk and lend support to this picture. In addition, folded \textit{RXTE} light curves seem to indicate complex changes in the geometry of the accretion regions on the surface of the neutron star, which may be indicative of an inhomogeneous density distribution in the circumstellar material causing a variable accretion rate onto the neutron star. Finally, the assumed inclination of the system and H$\alpha$ equivalent width measurements are used to make a simplistic estimate of the size of the circumstellar disk.
\end{abstract}


\section{Introduction}

\begin{figure}
 \includegraphics[width=60mm,angle=90]{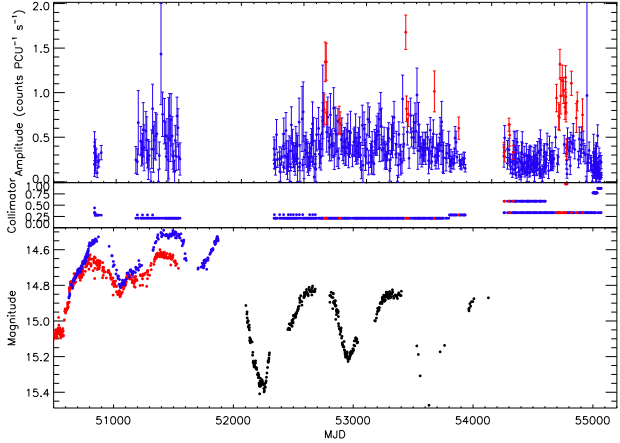}
  \caption{SXP6.85 X-ray and optical light curves. The top panel shows the PCA pulsed flux light curve; the middle panel shows the RXTE collimator response to SXP6.85 and the final panel shows the combined MACHO (red), OGLE II (blue) and OGLE III (black) light curves in the B (MACHO) and I (OGLE) photometric bands. The MACHO data have been arbitrarily normalised for viewing purposes.\label{fig:lightcurve}}
\end{figure}

Be/X-ray binaries are systems in which a neutron star orbits a massive, early type star that at some stage has shown evidence of line emission in the Balmer series. The neutron star accretes via interactions with an extended envelope of material in the equatorial plane of the Be star. These systems typically have wide, eccentric orbits meaning most X-ray outbursts are associated with the neutron star's passage through periastron. These Type I outbursts are generally in the luminosity range $10^{36}-10^{37}erg s^{-1}$ and last a few days. The less common Type II outbursts are brighter, $\gtrsim$\,$10^{37} erg s^{-1}$, and have no correlation with orbital phase. They can last from a few days to several months, depending on the size and state of the circumstellar disk.

The Small Magellanic Cloud (SMC) is known to house $\sim$\,60 such Be/X-ray binary systems \citep{coe08,cor08}, making up all but one of the known High Mass X-ray Binary (HMXB) population (SMC X-1 being the only known supergiant system). \citet{mcbride08} present spectral classification of the counterparts to a large fraction of these systems and show that the spectral distribution is consistent with that of the Galaxy, despite the lower metallicity environment.

The source that is the subject of this paper is the SMC Be/X-ray binary pulsar SXP6.85 = XTE J0103-728. It was first detected in 2003 by the \textit{Rossi X-ray Timing Explorer (RXTE)} as a 6.848 second pulsed X-ray source \citep{cor03}. Lomb-Scargle periodograms of the X-ray data reveal a 112.5 day period \citep{gal08}, although it is uncertain as to whether this modulation is the orbital period of the system or if it is driven by the interval between X-ray outbursts. The system was later detected in a 2006 \textit{XMM-Newton} observation at the position $01^{h}02^{m}53^{s}.1, -72^{\circ}44^{'}33^{''}.0$ (J2000.0). This detection led to the identification of a V=14.6 optical counterpart \citep{hab08}. Follow-up work by \citet{mcbride08} classified the counterpart as an O9.5V-B0V emission line star.

Previous analysis of MACHO red and blue data by \citet{kem08} reveals a $620 \pm 18$\,d optical modulation in which the source brightens by $\sim$\,0.5 magnitudes. Those authors also show that the source gets redder as it gets brighter. \citet{schmit07} analysed MACHO and OGLE II data of the source, finding an optical variation of $\sim$\,658 days. This is similar to the value found by \citet{kem08}, providing evidence for a quasi-periodic growth and decay of a circumstellar disk. \citet{schmit07} also find a possible period of 24.82 days in the OGLE II data. This period is closer to the expected orbital period of the system based on the Corbet diagram \citep{cor86} than the X-ray modulation found in \citet{gal08}. In this paper, we present long X-ray and optical lightcurves of SXP6.85 and analysis of X-ray timing and spectroscopic data taken by \textit{RXTE} and \textit{INTEGRAL}. Near-IR photometric data taken during the most recent outburst are presented and compared to near-IR data taken during quiescence to show the connection between optical and X-ray activity in this system. Ground based H$\alpha$ spectroscopic observations taken during the recent X-ray outburst are also presented and the equivalent width and assumed inclination are used as a simplistic measure of the size of the circumstellar disk.

\section{X-ray Data}

\subsection{\textit{RXTE}}

\begin{figure}
 \includegraphics[width=60mm,angle=90]{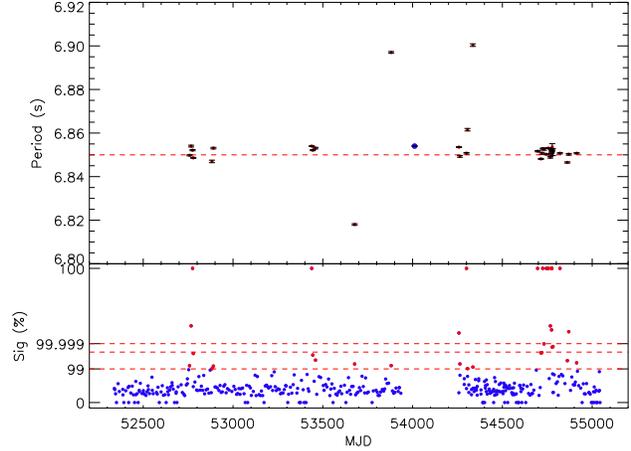}
  \caption{X-ray pulse period history of SXP6.85 and the significance of each detection according to a Lomb-Scargle periodogram. Only significant detections of the pulse period have been plotted; this corresponds with the 99 percent significance level (red points in bottom panel). The pulse period detected by \textit{XMM-Newton} in 2006 has been plotted alongside the \textit{RXTE} detections (blue point in upper panel; MJD 54011).\label{fig:period}}
\end{figure}

During the past 11 years, the SMC has been the subject of extensive monitoring \citep{laycock05,gal08} using the \textit{RXTE} Proportional Counter Array (PCA) (see \citet{jah06} for instrument details and calibration model). During this time, we have detected outbursts from SXP6.85 above the 99.99 percent significance level on four separate occasions. The full \textit{RXTE} light curve is shown in Fig. \ref{fig:lightcurve}. The red points on the upper two panels show \textit{RXTE} detections above the 99 per cent significance level. The PCA collimator response during the most recent outburst was 0.33, except for a period of $\sim$\,10 days in November 2008 in which an intense set of observations were carried out pointing directly at the source. These observations were in addition to weekly monitoring and were taken due to the unprecedented longevity of the outburst, which lasted for $\sim$\,20 weeks. The X-ray outbursts are shown in Fig. \ref{fig:lightcurve} between MJD 52700 and MJD 54900, with the detected spin period and detection significance plotted in Fig. \ref{fig:period}. Data reduction was performed using Heasoft v.6.6.2 analysis tools.

\begin{figure}
 \includegraphics[width=84mm]{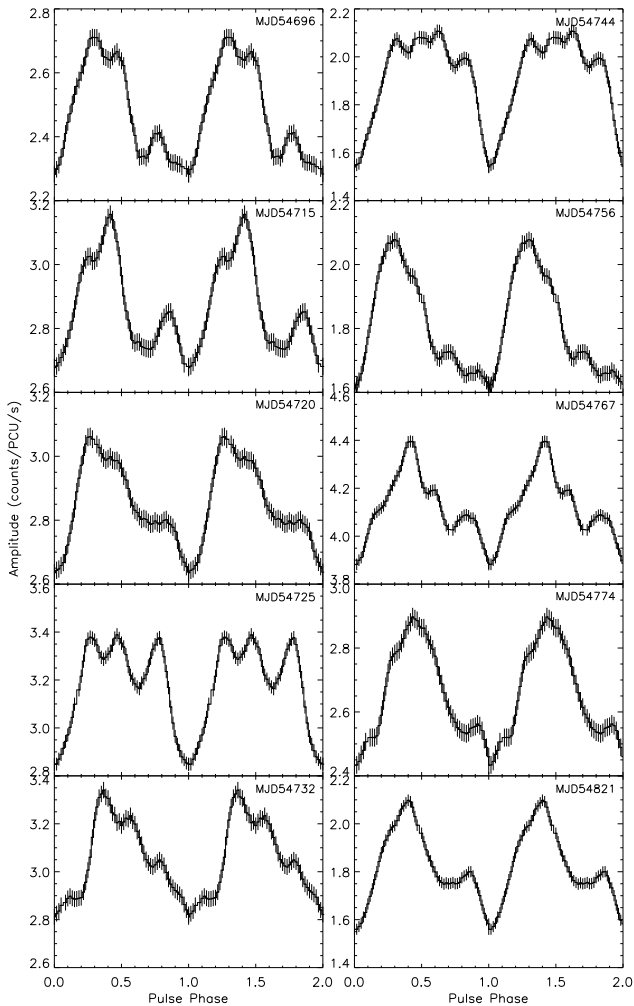}
  \caption{Pulse profiles of SXP6.85 during the most recent X-ray outburst (MJD = 54688 - 54830). The profiles are 3-10 keV light curves folded at the pulse period and have been smoothed and arbitrarily shifted in phase to align the minima in amplitude. Thus, they are included as a visual representation of the changes in the X-ray emission regions on the surface of the neutron star and do not contain any absolute information regarding the neutron star phase.\label{fig:test}}
\end{figure}

The timing and spectral analysis presented below is a result of \textit{RXTE} and \textit{INTEGRAL} data taken during the most recent epoch of pulsed X-ray emission. \textit{RXTE} observations were made approximately weekly between 2008 August 10 and 2008 December 30. According to the ephemeris in \citet{gal08}, the maximum in X-ray luminosity of the system is expected to occur on 2008 November 23 (MJD 54793). Although this date coincides with the detection, it is clear that the length and brightness indicate a Type II outburst and not a smaller Type I outburst usually associated with passage through periastron. Figure \ref{fig:test} shows \textit{RXTE} pulse profiles in the 3-10\,keV energy band for 10 observations throughout the outburst. The light curves were folded at the pulse period and arbitrarily phase shifted to align the point of minimum amplitude.

\subsection{\textit{INTEGRAL}}

\begin{figure}
 \includegraphics[width=60mm,angle=90]{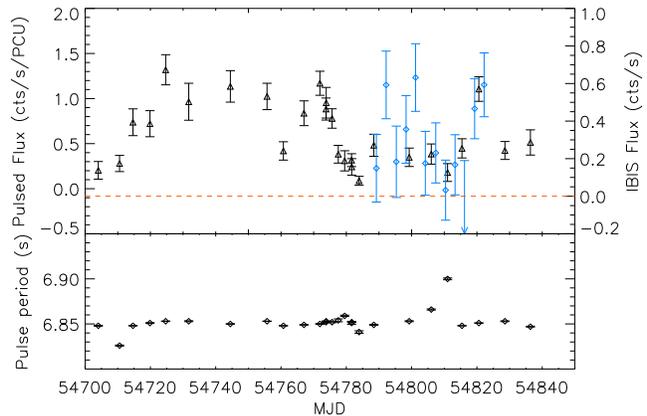}
  \caption{Combined \textit{RXTE} and \textit{INTEGRAL} light curve of SXP6.85. The black data points are the 3-10 keV pulsed flux detected with \textit{RXTE}, while the blue data points are the 15-35 keV total flux detected with \textit{INTEGRAL}. The pulse period in the lower panel is from timing analysis of \textit{RXTE} data. The dashed line corresponds to zero flux as detected by IBIS.\label{fig:integral}}
\end{figure}

The IBIS telescope aboard \textit{INTEGRAL}, which is optimised for an energy range of 15-200\,keV and has a field of view of 30$^{\circ} \times$ 30$^{\circ}$, is uniquely suited to observing large sky areas for point sources. As part of a key programme monitoring campaign on the SMC and 47 Tuc, \textit{INTEGRAL} observed the SMC and Magellanic Bridge for approximately 80\,ks per satellite revolution from 2008 November 11 to 2008 December 21. Individual pointings (science windows) were processed using the \textit{INTEGRAL} Offline Science Analysis v.7.0 (OSA) \citep{gold03} and were mosaiced into revolution sky maps using the weighted mean of the flux in the 15-35\,keV energy range. Figure \ref{fig:integral} shows the IBIS detections of SXP6.85 plotted over a section of the \textit{RXTE} pulsed flux light curve from Fig. \ref{fig:lightcurve}. These data represent the hard X-ray flux measured from SXP6.85. The brightest detection at 0.63 cts/s corresponds to a peak luminosity of 1.1$\times$10$^{37}$\,erg/s. The significance of the detection of the source once all science windows from the 33\,d of coverage had been taken into account was 6.4 sigma. We used the four brightest detections to make an X-ray spectrum to compare with the softer \textit{RXTE} spectrum. The \textit{RXTE} spectrum was taken on MJD 54772, approximately 4 weeks before the \textit{INTEGRAL} data. Although this is not ideal for comparison, we are limited to the brightest detections with both satellites and to detections with \textit{RXTE} in which no other source was active.

\begin{figure}
 \includegraphics[width=55mm,angle=270]{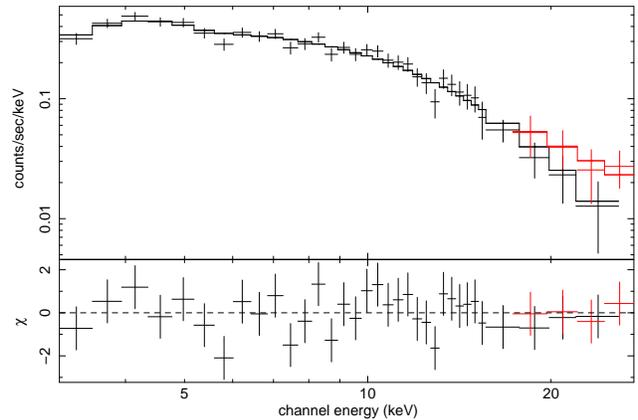}
  \caption{Typical X-ray spectrum of SXP6.85 extracted from an observation in which no other pulsating sources were in emission to minimise contamination effects. The harder IBIS spectrum (red) was fit simultaneously with the \textit{RXTE} spectrum (black) with a freely variable offset. The spectrum has been rebinned above 15\,keV due to low counts. See text for discussion of model fitted.\label{fig:xspec}}
\end{figure}

The \textit{RXTE} and \textit{INTEGRAL} spectra were fitted with an absorbed cutoff powerlaw model using XSPEC version 11.3.2 \citep{arnaud96}. The absorption was frozen at $6\times10^{20}$\,cm$^{-2}$ \citep{dic90}. The power law had a photon index of 0.8$\pm$0.2 with a cutoff at 14.6$\pm$9.6\,keV. The two spectra were fit with this model simultaneously with an offset allowed to vary during the fit. A \textit{$\chi^{2}_{\nu}$} of 0.75 (33 d.o.f) indicates a good fit to the data (see Fig. \ref{fig:xspec}). This analysis reveals a peak X-ray luminosity of $\sim$\,7$\times10^{36}$ erg\,s$^{-1}$ in the 3-10 keV band, assuming a distance to the SMC of 60\,kpc. Only \textit{RXTE} spectra in which there was no other detected pulsed X-ray emission were used to calculate $L_{X}$. We find that this value varies by a factor of two during the outburst. Analysis of \textit{XMM} data by \citet{hab08} show the source at a higher luminosity, $L_{X}$\,=\,$2\times10^{37}$ erg\,s$^{-1}$ in the 0.2-10 keV band, and showing variations of a factor of two in intensity on time scales of ten minutes.

\section{Optical and IR Data}

\subsection{OGLE III}

The Optical Gravitational Lensing Experiment (OGLE) project has been steadily monitoring millions of stars in the Magellanic Clouds for the past 12 years. The 1.3m Warsaw telescope at Las Campanas Observatory, Chile, takes photometric images in the I band, which now make up a decade long database that includes most of the HMXB systems in the SMC (see \citet{uda97} and \citet{syzm05} for more details on the instrument and catalogue). The bottom panel in Fig. \ref{fig:lightcurve} shows the OGLE II, OGLE III and MACHO \textit{blue} light curves of SXP6.85 (blue, black and red points respectively). The MACHO data have been arbitrarily normalised for viewing purposes.

It is clear that the long period oscillations described in \citet{kem08} are also present in the OGLE data, indicating a quasi-periodic brightening that is intrinsic to the Be star; most likely the growth and decay of the circumstellar disk. Although the OGLE III light curve does not extend up to the most recent \textit{RXTE} detection (this particular source falls on an OGLE chip-gap, thereby needing a different reduction process to the normal pipeline), it is apparent that the most luminous X-ray outbursts seem to coincide with the epochs of maximum optical brightness. This again suggests that the X-ray detections are Type II outbursts, being independent of binary phase. However, it is also apparent that there are 2 detections which fall at times of optical minima (Fig. \ref{fig:lightcurve} \& Fig. \ref{fig:period}). Thus, we cannot rule out there also being accretion onto the neutron star during times of a reduced circumstellar disk. We also note that the \textit{XMM} detection is consistent with an optical maximum.

\begin{table*}
 \centering
% \begin{minipage}{140mm}
%  \arraystretch{140mm}
  \caption{IR data of the counterpart to SXP6.85. \label{ta:irsf}}
  \begin{tabular}{@{}llrrrr@{}}
  \hline
   Catalogue & Reference ID & Date (MJD) & J & H & $K_{s}$\\
  \hline
   2MASS$^{1}$ & 01025333-7244351 & 51034.784 & $14.83\pm0.04$ & $14.71\pm0.05$ & $14.78\pm0.11$\\
   SIRIUS$^{2}$ & 01025331-7244351 & 52894.036 & $14.82\pm0.02$ & $14.74\pm0.02$ & $14.57\pm0.02$\\
  \hline  
   Telescope & & Date (MJD) & J & H & K$_{s}$\\
  \hline
   IRSF & & 54453 & $14.84\pm0.02$ & $14.77\pm0.02$ & $14.67\pm0.03$\\
        & & 54809 & $14.53\pm0.02$ & $14.46\pm0.02$ & $14.36\pm0.04$\\
  \hline
  \end{tabular}\\ 
 \begin{flushleft}
   $^{1}$\citet{skr06}, $^{2}$\citet{kato07}.
\end{flushleft}
%\end{minipage}
\end{table*}

\subsection{IRSF}

The 1.4m Infrared Survey Facility (IRSF) telescope at the \textit{South African Astronomical Observatory (SAAO)} is a Japanese built telescope designed specifically to take photometric data in the J, H \& $K_{s}$ bands. The SIRIUS (Simultaneous three-colour InfraRed Imager for Unbiased Survey) camera attached consists of three 1024 $\times$ 1024 pixel HAWAII arrays which take simultaneous images in the three bands \citep{nag99}. The pixel scale of the chip is 0.45 arcsec per pixel, yielding a $7'.7\times7'.7$ field of view.

Data reduction was performed using the dedicated SIRIUS pipeline based on the National Optical Astronomy Observatory's (NOAO) IRAF software package. The pipeline was provided by Yasushi Nakajima at Nagoya University, Japan. This performs the necessary dark subtraction, flat fielding, sky subtraction and recombines the dithered images. We generally used 15 dithers at 25s each per target, so a total exposure time of between 350 and 400s for each source. This depth is sufficient to detect sources down to $K_{s}$=17. Photometry on the reduced data was done using the NOAO/DAOPHOT package in IRAF to extract the J, H \& $K_{s}$ band magnitudes. The catalogue by \citet{kato07} was used to calibrate the data.

Table \ref{ta:irsf} shows the J, H \& $K_{s}$ magnitudes of the counterpart to SXP6.85 from both the 2MASS and SIRIUS catalogues \citep{skr06,kato07}. The IR position of the source is agreed by both catalogues to be $01^{h}02^{m}53^{s}.3, -72^{\circ}44^{'}35^{''}.1$ (J2000.0). Below the catalogue values are the data from two observing runs at the IRSF observatory during December 2007 and December 2008. Immediately obvious is the increase in near-IR brightness of the source between these two dates, suggesting an increase in the size of the circumstellar disk building up to the most recent X-ray outburst.

\subsection{Optical Spectroscopy}

\begin{figure}
 \includegraphics[width=60mm,angle=90]{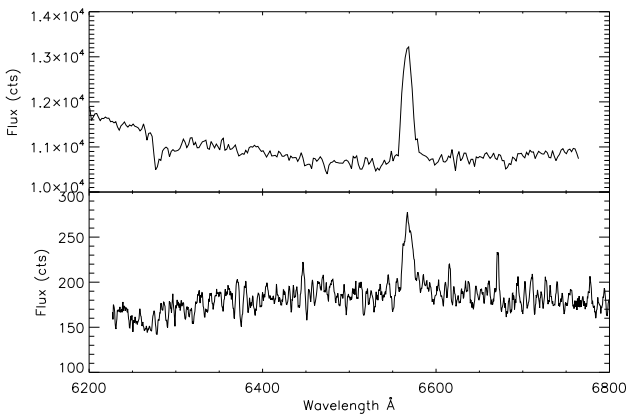}
  \caption{Top panel: H$\alpha$ spectrum of SXP6.85 taken with the ESO 3.6-m telescope on 2007 September 17. Bottom panel: H$\alpha$ spectrum of SXP6.85 taken with the SAAO 1.9-m telescope on 2008 December 16.\label{fig:spec}}
\end{figure}

In order to more fully understand the circumstellar disk of the system, H$\alpha$ spectroscopy was obtained simultaneously with the near-IR photometry using the 1.9-m telescope, Sutherland Observatory, South Africa. The spectrograph was used with the SITe detector and a 1200 l/mm grating. The dispersion was 1.0\AA{}/pixel and the SNR $\sim$\,10. A total of 9 spectra were taken throughout the two weeks beginning 2008 December 3. This was to check for any short-term variations in the H$\alpha$ Equivalent Width (W$_{eq}$) during the ongoing Type II X-ray outburst. Details of the observations made and the measured W$_{eq}$ are shown in Table \ref{ta:halpha}. Data reduction was performed using standard IRAF routines. In addition, another H$\alpha$ spectrum was taken in 2007 September 17 by the European Southern Observatory (ESO) 3.6-m telescope, La Silla, Chile, shortly after the optical counterpart had been verified. The EFOSC2 faint object spectrograph mounted at the Cassegrain focus was used with grism no:18 with lines ruled at 600/mm and which covered the wavelength range 4770-6770\,\AA{}. Using a slit width of 1$^{''}$ resulted in spectra at a resolution of 5\,\AA{}. The data reduction was performed using standard IRAF routines. The 3.6-m spectrum and one of the 1.9-m spectra are shown in the top and bottom panels of Fig. \ref{fig:spec} respectively. All of the spectra described have been smoothed with a boxcar average of either 5 or 7\,\AA{}.

\begin{figure}
 \includegraphics[width=60mm,angle=90]{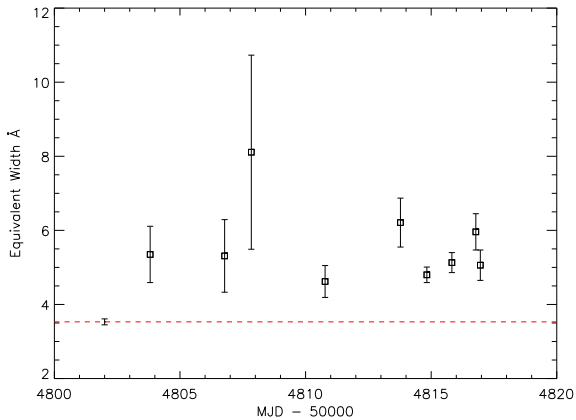}
  \caption{H$\alpha$ W$_{eq}$ measurements of SXP6.85 taken at the SAAO 1.9-m telescope during December 2008. The large error bars associated with the first three values are due to a lower SNR from shorter observations. The horizontal dashed line represents the H$\alpha$ W$_{eq}$ of the source during the September 2007 observation. The associated error in this case is much lower due to the higher SNR that can be obtained at the ESO 3.6-m telescope.\label{fig:eqw}}
\end{figure}

It is clear that there is H$\alpha$ in emission at both epochs, showing the existence of a circumstellar disk. The higher SNR spectrum in the top panel of Fig. \ref{fig:spec} shows a single peaked emission line structure. Using the standard rotation model put forward by \citet{str30}, we suggest this system is of low inclination, with the disk near face-on to our line-of-sight. W$_{eq}$ measurements suggest that the H$\alpha$ line flux has grown in the time since the 2007 observation by over 50 percent. This along with the IR data presented in $\S$3.2 provides strong evidence that the disk has grown. Figure \ref{fig:eqw} shows the 9 W$_{eq}$ measurements from the data taken with the 1.9-m telescope in December 2008. The first three data points have large associated errors due to the shorter exposure times used (1000s as opposed to 1500s for the remaining exposures). Even so, it is clear that, within these errors, there is no evidence for any variation in H$\alpha$ emission during the two weeks.

\begin{table}
 %\centering
 \begin{minipage}{170mm}
  %\arraystretch{140mm}
  \caption{H$\alpha$ W$_{eq}$ measurements of SXP6.85.\label{ta:halpha}}
  \begin{tabular}{@{}lrrr@{}}
  \hline
   Telescope & Date & Exposure time (s) & H$\alpha$ W$_{eq}$ (\AA{})\\
  \hline
   ESO 3.6-m & 2007-09-17 & 1000 & 3.53\,$\pm$\,0.08\\
   SAAO 1.9-m & 2008-12-03 & 1000 & 5.35\,$\pm$\,0.76\\
              & 2008-12-06 & 1000 & 5.31\,$\pm$\,0.98\\
              & 2008-12-07 & 1500 & 8.11\,$\pm$\,2.62\\
              & 2008-12-10 & 1500 & 4.62\,$\pm$\,0.43\\
              & 2008-12-13 & 1500 & 6.21\,$\pm$\,0.66\\
              & 2008-12-14 & 1500 & 4.80\,$\pm$\,0.21\\
              & 2008-12-15 & 1500 & 5.13\,$\pm$\,0.27\\
              & 2008-12-16 & 1500 & 5.96\,$\pm$\,0.49\\
              & 2008-12-16 & 1500 & 5.06\,$\pm$\,0.41\\
  \hline
  \end{tabular}\\
 \end{minipage}
 %\begin{flushleft}
%\end{flushleft}
\end{table}

\section{Discussion}

The optical light curve of SXP6.85 displays long-term, quasi-periodic variations which are most likely associated with the growth and decay of the Be star's circumstellar disk. Changes in the (B-R) colour support this idea. By comparison with the X-ray light curve we find that at least two of the X-ray outbursts are consistent with optical maximum. This has also been observed in other such systems (see Fig.2 of \citet{schurch09}).

The X-ray light curve shows no periodic behaviour that can be attributed to a binary period. All the detected outbursts seem to be driven by the intrinsic behavior of the Be star. We note that only small fluctuations in the pulse period are observed and that these show no correlation with time. Of greater interest is the pulse period history of the most recent Type II outburst (MJD 54688 - MJD 54830), which shows no general spin-up trend. This is an unusual observation as most HMXB systems show spin up of the neutron star during Type II outbursts due to the high accretion torques present.

Although we were unable to gain any quantitative handle of how the pulsed emission changed during the outburst, the pulse profiles show very complex double and triple peaked structure (see Fig. \ref{fig:test}). This tells us much about the geometry of the system and of the beam emission regions: We see both poles from the pulsar, with the emission from the poles asymmetrical and, except for the profiles corresponding to MJD = 54725 and 54744 in Fig. \ref{fig:test}, always favouring one pole. This implies accretion onto the poles is asymmetric. The variability in the shape and width of the most prominent peak suggests we are seeing both fan and pencil beam emission. However, it has been observed that pulse profiles are very messy at low energies, with the possibility that the absorption is varying with neutron star spin phase, so it is difficult to confirm that we are indeed seeing both fan and pencil beams in these energy ranges. \citet{bas75} theorise that pencil beams should be dominant under low accretion rates (hence low luminosities) and that fan beams should dominate at high accretion rates, as the radiation is strongly absorbed by the infalling material and can largely only be emitted sideways. However, from the profiles presented here there seems to be no clear dependence of the profile shape on flux.

\begin{figure}
 \includegraphics[width=60mm,angle=90]{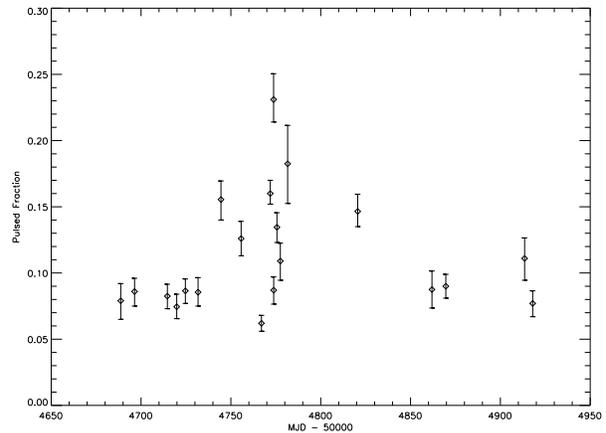}
  \caption{Pulsed fraction as a function of time in the 3-10\,keV band for the most recent SXP6.85 outburst. The data show variations of a factor of two in pulsed fraction on the timescale of days. The definition of pulsed fraction used in this work is given in the text.\label{fig:pulse_frac}}
\end{figure}

The pulsed fraction calculated from the folded light curves is shown in Fig. \ref{fig:pulse_frac}. The formula used in these calculations is given in Eqn. \ref{equ:pf}. We show that there is considerable variation in the measured pulsed fraction throughout the outburst, with the source appearing to increase and then return to some base level.

\begin{equation}
 {\rm Pulsed Fraction} = \frac{(F_{max} - F_{min})}{(F_{max} + F_{min})}
  \label{equ:pf}
\end{equation}

\noindent We note that the pulsed fraction is only correct if SXP6.85 was the only source detected in that observation. If other sources were detected, then the base emission includes emission from those objects and as such the pulsed fraction is only the lower limit of the actual value. We find a marginal correlation between source intensity and pulsed fraction, which could be attributed to the fractional change in the contribution of quiescent sources in the field. Transient sources have been previously detected at low luminosities while in quiescence (i.e. without any pulsed emission). Therefore, although we can be sure there was no bright pulsed emission from any source in the field of view, we cannot be certain that there was no continuous emission. The values presented in Fig. \ref{fig:pulse_frac} are a factor of three lower than the value presented in \citet{hab08}, having taken the method of calculating the pulsed fraction into account. This implies either a significant contribution from background sources to the \textit{RXTE} spectrum or that there has been a dramatic change of pulsed emission between the two observations. Those authors also find that SXP6.85 has one of the hardest spectra of known HMXB systems with a photon index of $\sim$\,0.4. This also suggests that there is contamination from other sources in the \textit{RXTE} spectrum, softening the powerlaw photon index to the value of 0.8 found in this study.

It has been observed in other Be/X-ray binary systems that the (J-K) colour often remains constant for long periods of time despite dramatic changes in near-IR magnitude (e.g. \citet{coe06}). SXP6.85 also exhibits this behavior as can be seen in Table \ref{ta:irsf}. The brightness of the system has increased by almost one third of a magnitude in the year between the two IRSF observations. However, the (J-K) colour has remained unchanged at 0.17. This is indicative of the temperature of the circumstellar disk remaining constant, whilst the increased brightness indicates the surface area of the disk has increased. It is possible to make a rough guess at the size of the circumstellar disk using the H$\alpha$ W$_{eq}$ measurements and making some assumptions on inclination and optical depth. Using the models of \citet{grund07}, we estimate the disk half-width at half-maximum (HWHM) radius to be approximately 3\,R$_{*}$ on 2008 December 16 and approximately 2.7\,R$_{*}$ on 2007 September 17, an increase of over 11 percent in the year leading up to the recent Type II outburst. In this calculation we do not take into account any truncation caused by the neutron star and we make some approximations regarding the binary system: the effective temperature of the star is 30000\,K, the inclination is 10$^{\circ}$, the optically thin outer boundary of the disk is at 50\,R$_{*}$ and the disk continuum dilution factor is 0.0033 \citep{dachs88}. See \citet{gg06} and \citet{grund07} for more detail on the model input parameters. The key thing to note is that varying the disk outer boundary by factors of 2-3 changes the calculated HWHM radius by approximately 10 percent as the model assumes a small contribution to the H$\alpha$ flux from the optically thin part of the disk. Thus, the relative growth of the disk is set by the model despite the uncertainty in the actual radius. We also note that the 2008 measurement was made after SXP6.85 had been in outburst for more than 16 weeks, and as such the disk may have been significantly larger at the start of the outburst than the value estimated here. If we consider the IR magnitudes taken in December 2007 and December 2008, this yields a fractional increase in brightness of 1.33; a factor of three higher than the prediction made using the equivalent widths. The OGLE light curve shows variations on the order of 0.7 magnitudes over the entire epoch of coverage, corresponding to a factor of 1.9 in brightness (here we assume that the change in flux is proportional to the change in surface area of the disk, which is reasonable for an optically thick disk). If we assume that the recent X-ray activity occured during an optical maximum and that the disk HWHM at this time was $\sim$\,3\,R$_{*}$, this implies the minimum, or quiescent, level of the disk is at $\sim$\,1.5\,R$_{*}$. Clearly these are rough estimates based on some simplistic approximations. More detailed modelling is encouraged to reveal a more accurate description of the circumstellar disk in this system.

\section{Conclusions}

The X-ray and optical data presented here are evidence that SXP6.85 is indeed a Be/X-ray binary system. The X-ray light curve shows the transient nature of the source, whilst the optical and near-IR data suggest there are quasi-periodic modulations that are intrinsic to the Be star that are likely to be associated with the extended envelope of matter in the equatorial plane of the companion. H$\alpha$ flux measurements show that an extended disk is indeed present in this system, with the shape of the emission line suggesting low inclination to the observer. The small variation in the spin period is also evidence of a low inclination system, assuming the circumstellar disk and the neutron star orbit lie in the same plane. W$_{eq}$ measurements of the H$\alpha$ line reveal a growth in the emission of H$\alpha$ photons from the system which is most likely a result of an increase in the size of the circumstellar disk; an argument that is supported by the increase in optical luminosity that culminated in the most recent epoch of pulsed X-ray emission from the neutron star. Folded X-ray light curves reveal the geometry of the emission region on the surface of the neutron star and show the highly variable nature of the accretion region. Finally, a simple estimate of the radius of the circumstellar disk was made, quantifying the size of the disk during periods of X-ray activity and during quiescence.

\section*{Acknowledgements}

This publication makes use of data products from the Two Micron All Sky Survey, which is a joint project of the University of Massachusetts and the Infrared Processing and Analysis Center/California Institute of Technology, funded by the National Aeronautics and Space Administration and the National Science Foundation. This paper utilises public domain data originally obtained by the MACHO Project, whose work was performed under the joint auspices of the U.S. Department of Energy, National Nuclear Security Administration by the University of California, Lawrence Livermore National Laboratory under contract No. W-7405-Eng-48, the National Science Foundation through the Center for Particle Astrophysics of the University of California under cooperative agreement AST-8809616, and the Mount Stromlo and Siding Spring Observatory, part of the Australian National University. The OGLE project is partially supported by the Polish MNiSW grant N20303032/4275. LJT wishes to thank the IRSF/SIRIUS team for providing the reduction pipeline for the IR photometry and the University of Southampton, whose support has made this research possible. We would like to thank the anonymous referee for their helpful and constructive comments.

\bsp

\label{lastpage}

\end{document}